\documentclass[%
 twocolumn,
superscriptaddress,
 amsmath,amssymb,
 aps,
floatfix,
longbibliography,
]{revtex4}

\usepackage{graphicx}
\usepackage{dcolumn}
\usepackage{bm}
\usepackage{booktabs}
\usepackage{graphicx}
\usepackage{dcolumn}
\usepackage{bm}
\usepackage{nowidow}
\usepackage{tabularx}
\usepackage{siunitx}
\usepackage[usenames,dvipsnames]{xcolor}
\usepackage[colorlinks=true, allcolors=black]{hyperref}
\usepackage{tikz}
\usepackage{times}
\usepackage{hyperref}

\hypersetup{
    citecolor = MidnightBlue,
    linkcolor = RedOrange
}

\usepackage{natbib}
\setcitestyle{square, comma, numbers,sort&compress, super}
\begin{document}

\newcommand{\YF}[1]{\textcolor{purple}{\,#1}}
\newcommand{\RS}[1]{\textcolor{magenta}{\,#1}}
\newcommand{\SH}[1]{\textcolor{violet}{\,#1}}
\newcommand{\LI}[1]{\textcolor{olive}{\,#1}}
\newcommand{\mk}[1]{{\color{blue}#1}}
\newcommand{\ED}[1]{{\color{cyan}#1}}
\newcommand{\DG}[1]{{\color{orange}#1}}

\preprint{APS/123-QED}

\title{Characterizing hydrogel behavior under compression with gel-freezing osmometry}

\author{Yanxia Feng}
\affiliation{Department of Materials, ETH Z\"{u}rich, 8093 Z\"{u}rich, Switzerland.}%

\author{Dominic Gerber}
\affiliation{Department of Materials, ETH Z\"{u}rich, 8093 Z\"{u}rich, Switzerland.}%

\author{Stefanie Heyden}
\affiliation{Department of Civil, Environmental and Geomatic Engineering, ETH Z\"{u}rich, 8093 Z\"{u}rich, Switzerland.}%

\author{Martin Kr\"{o}ger}
\affiliation{Department of Materials, ETH Z\"{u}rich, 8093 Z\"{u}rich, Switzerland.}%

\author{Eric R. Dufresne}
\affiliation{Department of Materials Science \& Engineering, Cornell University, Ithaca, USA}%
\affiliation{Laboratory of Atomic and Solid State Physics, Cornell University, Ithaca, USA}%

\author{Lucio Isa}
\affiliation{Department of Materials, ETH Z\"{u}rich, 8093 Z\"{u}rich, Switzerland.}%

\author{Robert W. Style}
\email[]{robert.style@mat.ethz.ch}
\affiliation{Department of Materials, ETH Z\"{u}rich, 8093 Z\"{u}rich, Switzerland.}%

\date{\today}

\begin{abstract}
Hydrogels are particularly versatile materials that are widely found in both Nature and industry.
One key reason for this versatility is their high water content, which lets them dramatically change their volume and many of their mechanical properties -- often by orders of magnitude -- as they swell and dry out.
Currently, we lack techniques that can precisely characterize how these properties change with water content.
To overcome this challenge, here we develop Gel-Freezing Osmometry (GelFrO): an extension of freezing-point osmometry.
We show how GelFrO can measure a hydrogel's mechanical response to compression and osmotic pressure, while only using small, $O(100$~\textmu{}L$)$ samples.
Because the technique allows measurement of properties over an unusually wide range of water contents, it allows us to accurately test theoretical predictions.
We find simple, power-law behavior for both mechanical response to compression, and osmotic pressure, while these are not well-captured by classical Flory-Huggins theory. 
We interpret this power-law behavior as a hallmark of a microscopic fractal structure of the gel's polymer network, and propose a simple way to connect the gel's fractal dimension to its mechanical and osmotic properties.
This connection is supported by observations of hydrogel microstructures using small-angle x-ray scattering.
Finally, our results motivate us to propose an updated constitutive model describing hydrogel swelling, and mechanical response.

\end{abstract}

\maketitle

\section{Introduction}

Polymer gels are almost unique in their ability to swell or shrink by tremendous amounts, greatly changing their properties, while maintaining their elastic nature.
Typically, solvent-swollen gels are soft, stretchable and permeable to transport of small molecules, while dry gels are stiff and impermeable, and tend to be glassy \cite{delavoipiere2016poroelastic,gehrke1997factors}.
This ability to completely change material characteristics makes gels particularly versatile materials.
This is especially true of hydrogels, which have the additional benefit of being generally biocompatible, due to their aqueous nature \cite{tibbitt2009hydrogels}.
In biology, hydrogels make up a large portion of soft tissue, and nature utilizes their swelling ability for applications ranging from controlling plant-tissue stiffness, to swelling-induced actuators, membranes, and barrier coatings like skin \cite{forterre2013slow,wu2024generating,na2022hydrogel,yuk2016skin}.
Beyond this, hydrogels are finding increasing industrial and scientific usage.
For example, they are well known as super-absorbent materials \cite{brannon2012absorbent,webber2023linear}, and have found extensive use as drug-delivery agents \cite{peppas2006hydrogels}, and in diverse applications including wound dressing \cite{mirani2017advanced}, agriculture \cite{liu2022environmentally}, and enhanced oil-recovery \cite{pu2017development}.
Hydrogels can also be made responsive to triggers such as light, pH and temperature \cite{shibayama2005volume}.
Thus, recent work has focused on making materials like stimuli-responsive membranes and actuators, and active materials \cite{moser2022hydroelastomers,jeon2017shape,sydney2016biomimetic,van2023self}.

To work with hydrogels capable of huge changes in volume, we need techniques for characterizing hydrogel properties over a wide range of compressions.
At a minimum, we need to know their mechanical constitutive behavior, and swelling properties (e.g. osmotic pressure).
However, these are surprisingly challenging to measure.
One approach is to perform `drained' compression tests, which let solvent drain out of the hydrogel, while measuring stress and strain \cite{kazimierska2015determination,bhattacharyya2020hydrogel,li2012experimental}.
This method works well for chemically-crosslinked gels, but requires large samples: at least $O$(mL) when using a typical rheometer.
The tests also usually take hours to days to perform, as, at each step, enough time must be given for the solvent to fully drain out of the sample \cite{hu2010using}.
Alternatively, a gel's osmotic pressure can be measured with techniques such as membrane- \cite{yasuda2020universal,mallam1989scattering} or vapor-pressure osmometry \cite{gao2021scaling}.
However, these also have associated challenges.
For example, membrane osmometry is limited to osmotic pressures of less than 5~MPa \cite{straub2014raising}, while vapor-pressure osmometry suffers from difficulties in accurate measurement of a solvent's vapor pressure \cite{madden2022comparison}. 
As with compression testing, both of these techniques require long experiments at the scale of hours, due to the need for lengthy equilibration times \cite{madden2022comparison}.

As a result of these challenges, there is very little data characterizing gel behavior over wide ranges of compression or swelling \cite{li2012experimental,bhattacharyya2020hydrogel}.
Here, we show how controlled freezing of hydrogels in the form of Gel Freezing Osmometry (GelFrO) offers a convenient technique for performing this characterization.
In particular, we extract both the large-strain constitutive response, and the osmotic pressure of a hydrogel, using $O(100$~\textmu{}L$)$  gel samples.
These samples drain rapidly in response to freezing, so each data point takes only a few minutes to collect.
Both osmotic pressure and constitutive response take power-law forms.
We suggest that this is due to hydrogels having a fractal microstructure, and use the results to derive a new constitutive model.

\section{Characterizing ice/hydrogel equilibrium}

GelFrO is essentially an extension of the common technique of freezing-point osmometry (FPO), which is widely used to measure solution properties of aqueous systems.
In FPO, the freezing temperature of an aqueous solution is measured, and this is related to the solution's osmotic pressure, \textit{$\Pi$}, by
\begin{equation}
    \rho_w L \left(\frac{T_m-T}{T_m} \right)=\Pi.
    \label{eqn:fpd}
\end{equation}
Here, $\rho_w$ is the density of water, $L$ is the specific latent heat of fusion of ice, $T$ is the temperature and $T_m$ is the freezing temperature of pure water at atmospheric pressure \cite{worster2021colloidal,style2023generalized}.
Here, temperatures are measured in Kelvin.
While this works well for solutions, it cannot be straightforwardly applied to hydrogels, chiefly because the polymer network of the gel suppresses ice nucleation, and this will affect the result \cite{zhuo2021gels,fernandez2024elastic}.
Furthermore, Equation (\ref{eqn:fpd}) needs to be modified to include additional terms from the elasticity of the polymer gel (see below). 

\begin{figure}[htbp]
\includegraphics[width=\columnwidth]{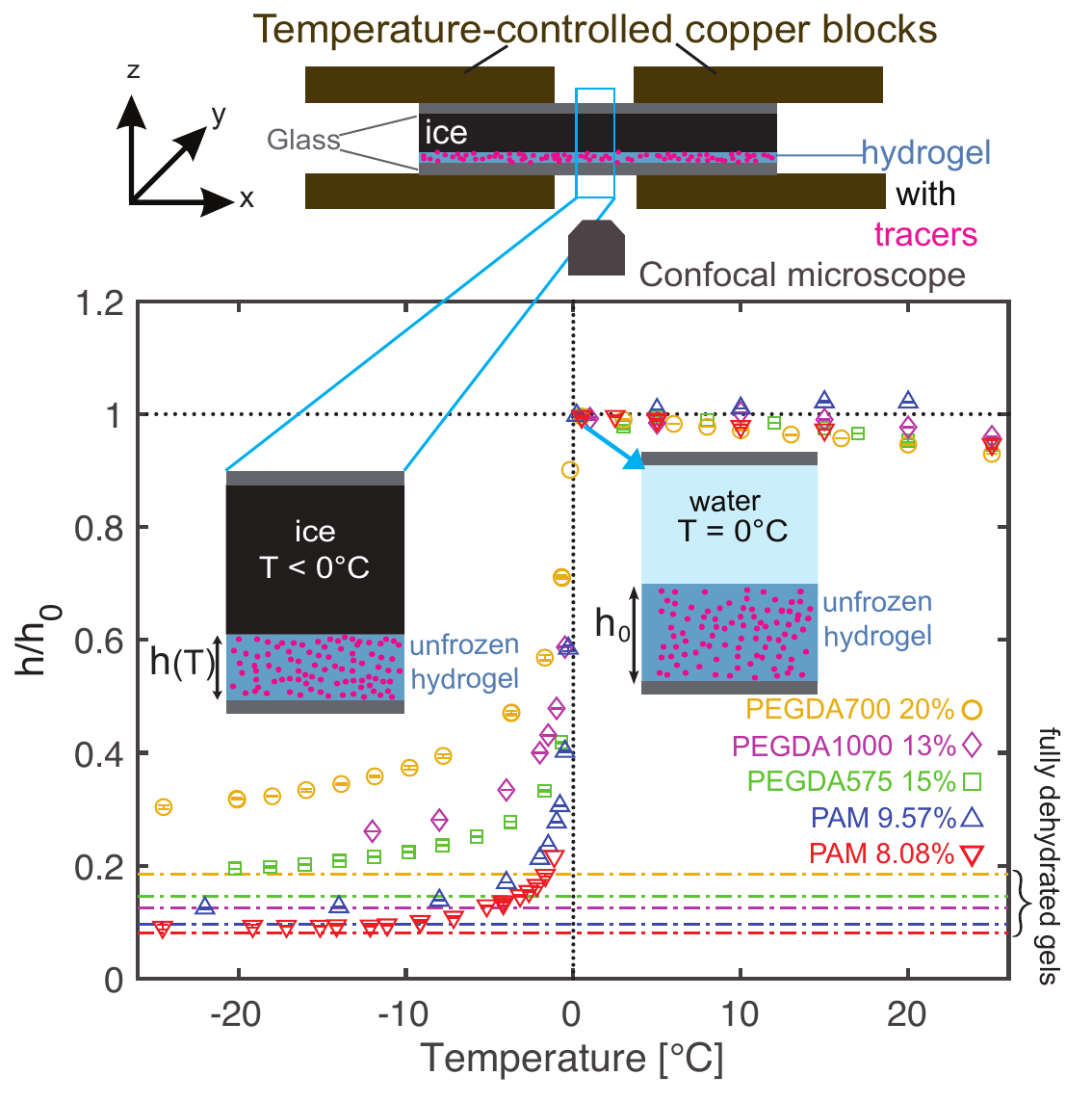}

\caption{The equilibrium relative thickness of hydrogel layers in contact with bulk ice/water. Top: schematic of the freezing stage. Bottom: change in hydrogel layer thickness as a function of temperature -- relative to the thickness at $T=0$~\textdegree{}C. Above 0~\textdegree{}C, hydrogels are in equilibrium with bulk water, and there is only a small dependence of film thickness on temperature. Below 0~\textdegree{}C, hydrogels are in contact with bulk ice, and shrink with increasing undercooling. Shrinkage is caused by cryosuction from the ice drawing water out of the hydrogel. Dash-dotted lines show the thickness of fully-dehydrated gels -- i.e. dry polymer. Temperature measurements are accurate to 0.05~\textdegree{}C, while error bars in $h/h_0$ are given by the standard deviations of three separate measurements on each sample at each temperature. Error bars are smaller than marker sizes.}

\label{fig:schematic}
\end{figure}

We re-imagine FPO for hydrogels by tracking the shrinkage of a gel layer in contact with ice, as the temperature reduces (Figure~\ref{fig:schematic}).
Ice at subzero temperatures essentially exerts a temperature-dependent osmotic pressure, or `cryosuction' on an adjacent piece of hydrogel, which sucks water out of the gel, causing it to shrink (the water in the hydrogel itself does not freeze, as ice is prevented from growing into the hydrogel by capillarity \cite{dash2006physics}).
Measuring this shrinkage allows us to characterize properties including large-deformation constitutive response and osmotic pressure. 

We demonstrate GelFrO with two of the most commonly-found charge-neutral, synthetic hydrogels: poly(ethylene glycol) diacrylate (PEGDA), and polyacrylamide (PAM).
Three different PEGDA hydrogels are made from PEGDA monomers with molecular weights of 575, 700, and 1000 g/mol with as-prepared polymer volume fractions of $\phi_0=$ 15, 20, 13\% respectively.
Two PAM hydrogels are made from a mixture of acrylamide monomer, and bisacrylamide crosslinker.
A softer gel is prepared with 8.01 vol\% monomer and 0.07 vol\% crosslinker (PAM 8.08\%), while a stiffer gel is made with 9.18 vol\% monomer and 0.39 vol\% crosslinker (PAM 9.57\%).
All the gels are bonded to a glass cell using a silane coupling agent (see Materials and Methods).
This choice of hydrogels provides materials with a range of different polymer volume fractions and mechanical properties.
Table~\ref{tab1} shows $\phi_0$, and the drained Young's modulus, $E_d$ for all the hydrogels.
$E_d$ was measured via indentation of bulk samples, for all the materials except PEGDA1000.
The latter was considered too expensive to make bulk samples for testing.

\begin{table*}[ht]

\caption{Hydrogel properties and parameters from power law fitting on osmotic pressure data and mass fractal fitting on SAXS data. $E_d$ was not measured for PEGDA1000. See the supplement for details of how $R_g$ is calculated.}

\begin{tabular}{l@{\qquad\quad}c@{\;\;\;}c@{\;\;\;}c@{\;\;\;}|@{\quad}c@{\;\;\;}c@{\;\;\;}|@{\quad}c@{\;\;\;}c}
\hline\hline
 & \multicolumn{3}{c}{Hydrogel properties} & \multicolumn{2}{c}{$\Pi$ fitting} & \multicolumn{2}{c}{SAXS fitting}\\
   Hydrogel & $\phi_0$ & $E_d$~[kPa] & $R_g$~[nm] &  $A_o$ [-]  & $d_o$ [-] & $d_s$ [-] & $ \xi$~[nm] \\
   \hline
  PEGDA575 & $15\%$ & $189 \pm 19$ & 0.99 & $1.00\pm0.11$ & $2.41\pm0.04$ & $2.81\pm0.00$ & 53.8 $\pm$ 5.6 \\
  PEGDA700 & $20\%$ & $932 \pm 156$  & 1.09 & $2.52\pm0.24$ & $2.35\pm0.02$ & $2.46\pm0.01$ & 27.6 $\pm$ 1.4\\
  PEGDA1000 & $13\%$ & --- & 1.30 & $3.16\pm1.51$ & $2.32\pm0.09$ & $2.30\pm0.02$ & 26.2 $\pm$ 1.8\\
  PAM 8.08$\%$ & $8.08\%$ & 16 $ \pm $ 1  & 0.20 & $0.26\pm0.02$ & $1.96\pm0.06$ & $2.04\pm0.25$ & 2.02 $\pm$ 0.37\\
  PAM 9.57$\%$ & $9.57\%$ & $90 \pm 7$  & 0.20 & $0.28\pm0.04$ & $2.06\pm0.08$ & $1.61\pm0.05$ & 5.96 $\pm$ 0.55\\
  \hline\hline
\label{tab1}
\end{tabular}

\end{table*}

We perform freezing experiments using a temperature-controlled setup that fits on the stage of a confocal microscope \cite{gerber2022stress,gerber2023polycrystallinity,gerber_stage}, as shown schematically in Figure~\ref{fig:schematic}.
This allows us to apply uniform undercoolings to thin hydrogel samples, while simultaneously measuring their change in volume by tracking the movement of fluorescent, tracer nanoparticles embedded in the gels.
The hydrogels are fabricated as $\sim 150$ \textmu{}m thick layers bonded to the bottom, glass surface of a water-filled sample cell that fits into the freezing stage (Figure~\ref{fig:schematic}, see Materials and Methods for more details).
The top, glass surface of the cell is only loosely attached so that no confining stresses arise during the freezing process.

We initially equilibrate hydrogel layers with bulk water in the sample cell at room temperature.
At each subsequent step, we slowly reduce the temperature to a new temperature at a rate of 1~\textdegree{}C/min, and measure the equilibrium gel thickness, $h$.
In general, equilibrium is reached in $<$\ 5 mins of changing the stage temperature, as indicated by $h$ reaching a stable value (see Supplement).
We also check that the strain in the hydrogel is constant across its thickness, indicating uniform hydrogel shrinkage (see Supplement). 
Below $T_m=$ 0~\textdegree{}C, we nucleate ice in the bulk water (see Materials and Methods), and continue the process of gel/ice equilibration and thickness measurement across a range of different subzero temperatures.

The dependence of hydrogel layer thickness on temperature is shown in Figure~\ref{fig:schematic} for the different hydrogels.
These are given in terms of relative thickness changes $h/h_0$, where $h_0=h(T_m)$.
For most of the samples, there is a small expansion as the temperature reduces from room temperature to 0~\textdegree{}C. 
Immediately after ice formation (vertical dotted line), $h/h_0$ rapidly reduces, as the hydrogel is dehydrated by cryosuction from the ice \cite{yang2024dehydration}. 
Indeed, the hydrogels shrink more than 50\% in the first 5 degrees.
Upon further cooling, the thickness change slows as the gels approach their fully dehydrated limit (horizontal dash-dotted lines).
During this whole process, there is no measurable shrinkage of the gel in the $x$ and $y$ directions, due to the bonding to the underlying substrate (see Supplement).

\section{G\lowercase{el}F\lowercase{r}O as a drained compression test}

We can interpret our measurements of $h/h_0$ in two distinct ways.
Firstly, this test is analogous to a uniaxial, drained compression test on the hydrogel.
Thus, the results give us information about the large-strain constitutive properties of the gel.
Secondly, we can extract information about the gel's osmotic pressure as a function of polymer content.

To relate our measurements to a hydrogel's constitutive response, we use an expression for a hydrogel's total stress \cite{hong2008theory,bertrand2016dynamics}:
\begin{equation}
\sigma_{ij}=\sigma_{ij}^\mathrm{el}-\Pi_\mathrm{mix}(\phi)\delta_{ij}-\frac{\Delta\mu}{v_w}\delta_{ij},
\label{eqn:eff_stress}
\end{equation}
where $\sigma_{ij}^\mathrm{el}$ is the elastic stress in the polymer gel network. $\Pi_\mathrm{mix}(\phi)$ is the mixing osmotic pressure due to polymer/solvent interactions -- assumed to only depend on the polymer volume fraction, $\phi$. 
$\Delta\mu$ is the difference between the chemical potential of the hydrogel's water content and that of pure water at atmospheric pressure at the same temperature.
$v_w$ is the volume of a molecule of water, and $\Delta\mu/v_w$ can be thought of as the pore pressure of the water in the gel.
This expression follows from the common Frenkel-Flory-Rehner hypothesis, which assumes that the (generally nonlinear) contributions from network elasticity, mixing osmotic pressure, and solvent chemical potential are additive
\cite{mckenna1990swelling}.
We also assume that there are no additional solutes in the hydrogel.
In our system, $\sigma_{zz}=0$, as the gel layer is in mechanical equilibrium with the overlying, stress-free ice.
Furthermore, water in the gel is in equilibrium with the overlying ice, which implies that $\Delta\mu/v_w=-\rho_w L (T_m-T)/T_m$ (see Materials and Methods). 
Thus,
\begin{equation}
    \rho_w L \left( \frac{T_m-T}{T_m} \right)=\Pi_\mathrm{mix}(\phi) - \sigma^\mathrm{el}_{zz}.
    \label{eq:cryosuc_versus_gel_balance}
\end{equation}
This is a modified version of Equation~(\ref{eqn:fpd}) that includes the elasticity of the crosslinked gel.
It can be interpreted as the cryosuction from the ice (left side) being balanced by the mixing osmotic pressure and elastic stress coming from polymer chain compression (right side).

To connect this freezing experiment to an equivalent mechanical test, we consider the stress state in a uniaxial, drained compression test on the same sample.
A hydrogel sample is immersed in a water bath, and slowly compressed with a pressure $P_\mathrm{ext}$, while the water in the gel has time to flow out of the gel.
Then, the immersed gel is in equilibrium with the surrounding bath, so that $\Delta\mu=0$.
In this case, the $z$-component of Equation~(\ref{eqn:eff_stress}) becomes
\begin{equation}
        \sigma_{zz}=-P_{\mathrm{ext}}=\sigma_{zz}^\mathrm{el}-\Pi_\mathrm{mix}(\phi).
        \label{eqn:P_ext}
\end{equation}
By comparing the two equations above, we see that applying stress in a drained test is equivalent to equilibrating the sample with ice, when $P_{\mathrm{ext}}=\rho_w L (T_m-T)/T_m$.
Given that $\rho_w=998$~kg/m$^3$, $L=334$~kJ/kg and $T_m=273.15$~K, every degree of undercooling effectively exerts 1.2~MPa of compressional stress on the hydrogel. 

\begin{figure}[hpt]
\includegraphics[width=\columnwidth]{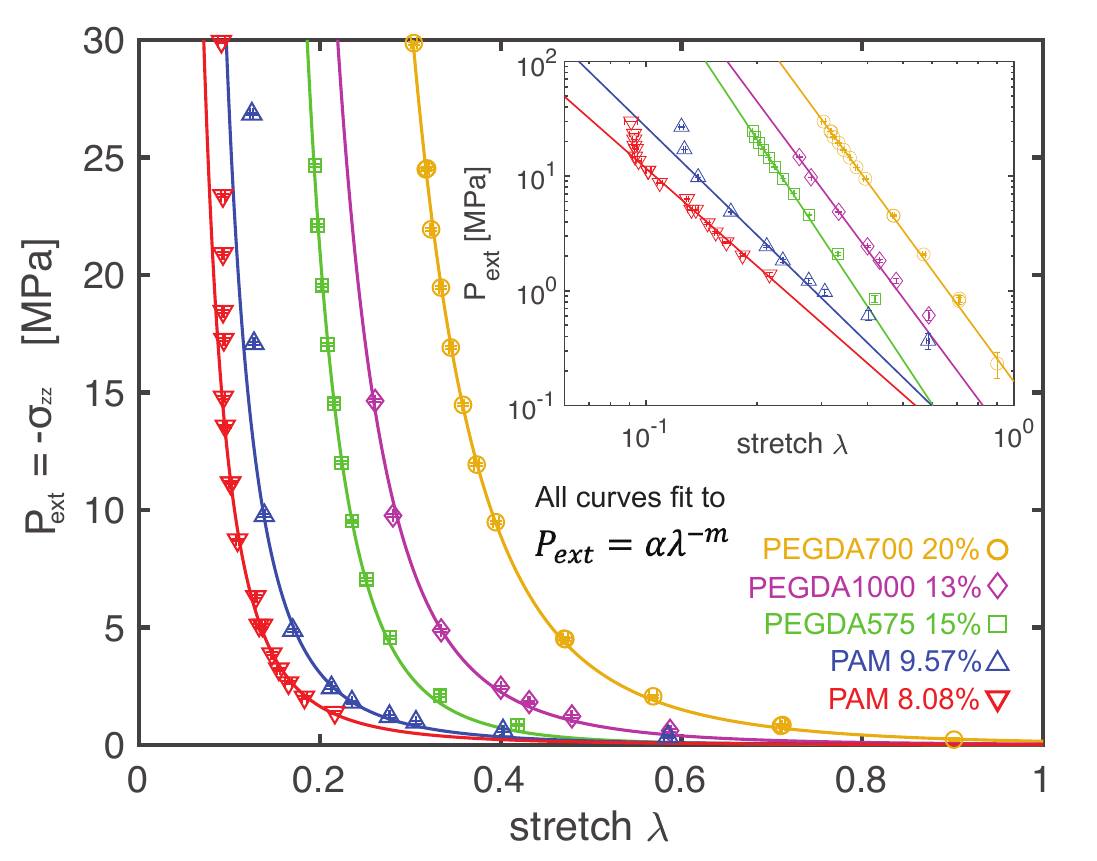}

\caption{The mechanical response of hydrogel samples in response to drained, uniaxial compression, with no lateral shrinkage. 
We obtain these results from the data in Figure~\ref{fig:schematic} by setting $P_\mathrm{ext}=\rho_w L (T_m-T)/T_m$ and $\lambda=h/h_0$.
Curves show power-law fits to the data.  
Inset: the same data on a log-log plot.}

\label{fig:stress-strain} 
\end{figure}

Based on these considerations, we can re-plot our data in Figure~\ref{fig:stress-strain} as the results of a drained compression test.
In particular, we plot $P_\mathrm{ext}$ as a function of the stretch in the sample, $\lambda=h/h_0$ in Figure~\ref{fig:stress-strain}.
The figure shows how the hydrogels dramatically strain-stiffen as they are compressed.
Close to the relaxed state ($\lambda=1$), the curves are all relatively flat, indicating a low tangent modulus (\emph{i.e.} stiffness).
Consistent with our mechanical indentation tests, the gels with larger $E_d$ (Table \ref{tab1}) have higher tangent moduli in this range.
As compression increases ($\lambda$ decreases), the curves become steeper, indicating significant stiffening.
Indeed, by comparing slopes, we see increases in stiffness from $<O(1~\mathrm{MPa})$ at $\lambda=1$, to $O(100~\mathrm{MPa})$ when compressed.
This fits with the intuition of hydrogels becoming much stiffer as they dry out.

Interestingly, the mechanical response of all the gels appears to be well-approximated by power laws.
This can be seen in the inset of Figure~\ref{fig:stress-strain}, which shows that all the data sets are approximately linear on a log-log plot.
This is especially true for the PEGDA hydrogels, while there are some deviations in the PAM gels at large deformation.

The results allow us to fit an empirical constitutive model to the hydrogel response over a large range of stretches.
This is by contrast to most other types of poroelastic mechanical tests, which do not allow such a fitting, as they only measure small-strain, linear response (e.g. \cite{hu2010using}).
For this fitting, we could use a simple power law: $P_\mathrm{ext}=\alpha \lambda^{-m}$.
However, this would not be stress-free at $\lambda=1$, as required.
Thus, we make the simplest possible extension to the power law, by adding a linear term:
\begin{equation}
    P_\mathrm{ext}=\alpha\left(\frac{1}{\lambda^m}-\lambda\right).
    \label{eqn:emp_fit}
\end{equation}
This empirical constitutive model captures both the power-law behavior at small $\lambda$, and has a stress-free state at $\lambda=0$.
Fits to this model are shown in the Supplement, and have a similar accuracy to the simple power-law fits shown in Figure~\ref{fig:stress-strain}.
Although Equation~(\ref{eqn:emp_fit}) is empirical, later, we shall see how it can be derived from a more general constitutive model for hydrogel behavior.

\section{Measuring gel osmotic pressure}

Our freezing results also permit measurement of $\Pi_\mathrm{mix}(\phi)$ (e.g. \cite{li2012experimental,cai2012equations}).
Equation~(\ref{eqn:P_ext}) tells us that the data in Figure~\ref{fig:stress-strain} represents the combined contribution of $\Pi_\mathrm{mix}$ and $\sigma^\mathrm{el}$ in the gel's polymer network.
However, the latter is generally negligible in our data, as $\sigma^\mathrm{el}\sim E_d(1-\lambda)$, and $E_d\ll P_\mathrm{ext}$ for the majority of our data.
This is especially true at large compressions.
Thus, to good approximation, $P_\mathrm{ext}\approx \Pi_\mathrm{mix}(\phi)$.

\begin{figure*}[htbp]
\includegraphics[width=\linewidth]{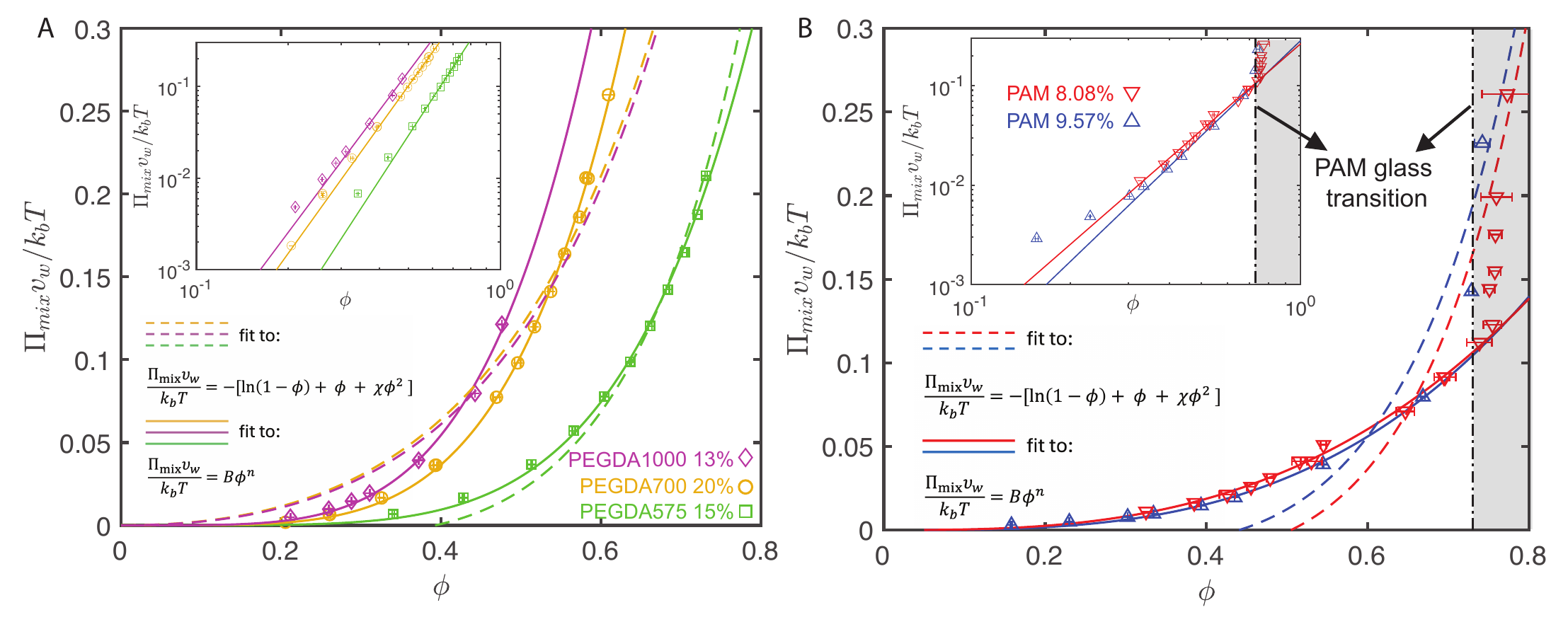}
\caption{Reduced mixing osmotic pressure as a function of polymer content for PEGDA gels (A), and PAM gels (B). Dashed curves show the best fits of the data to Flory-Huggins theory. Continuous curves show power-law fits. Insets show the same data on a log-log plot, with power-law fits. In (B), the grey area denotes the range of $\phi$ where PAM gels are expected to be glassy. Points in this area are not included in data fitting.}
\label{fig:gel_osmo} 
\end{figure*}

Based on these considerations, we convert Figure~\ref{fig:stress-strain} into a plot of osmotic pressure versus polymer content by setting $P_\mathrm{ext}=\Pi_\mathrm{mix}$ (Figure \ref{fig:gel_osmo}).
To calculate polymer volume fraction, we note that $\phi=h_p/h$ and $\phi_0=h_p/h_0$, where $h_p$ is the thickness of the completely dry hydrogel layer.
Combining this with the definition that $\lambda=h/h_0$ yields $\phi=\phi_0/\lambda$.
To aid later comparison with theoretical predictions, we plot the reduced osmotic pressure $\bar{\Pi}_\mathrm{mix}=\Pi_\mathrm{mix}v_w/k_bT$, where $k_b$ is Boltzmann's constant.
For all the gels, $\bar{\Pi}_\mathrm{mix}$ is small at low $\phi$, before increasing significantly as polymer fraction increases -- as expected for hydrophilic polymers.
Furthermore, comparing the three PEGDA hydrogels in Figure~\ref{fig:gel_osmo}A, we see that the higher the molecular weight of the PEGDA, the larger the osmotic pressure, and hence the more hydrophilic the gel is.
This agrees with previous results based on measurements of PEGDA's swelling properties in water \cite{malo2015heterogeneity}.

This technique allows us to measure osmotic pressure for a relatively wide range of polymer content.
Indeed, the PAM hydrogels in Figure~\ref{fig:gel_osmo}B reach over 70\% polymer at the coldest temperatures.
This is sufficiently large that some hydrogels are expected to go through a glass transition, $\phi_g$.
For PAM, we can estimate the glass transition as $\phi_g\approx 0.73$, based on previous work \cite{yuen1984glass}.
This is denoted in Figure~\ref{fig:gel_osmo}B by the grey areas.
The PEGDA samples are not expected to be glassy.

We can use our measurements of $\Pi_\mathrm{mix}(\phi)$ to test classical predictions from polymer physics.
In particular, Flory-Huggins theory for polymer/solvent mixtures forms the basis for the majority of theories describing hydrogel swelling, and predicts that \cite{flory1942thermodynamics,hong2008theory,bouklas2015nonlinear,bertrand2016dynamics}:
\begin{equation}
    \frac{\Pi_\mathrm{mix}v_w}{k_bT}=-\left[\ln(1-\phi)+\phi +\chi \phi^2\right].
    \label{eq:Flory_Huggins}
\end{equation}
Here, $\chi$ is the dimensionless monomer/solvent interaction parameter.
Note that this expression is based on assuming that polymerized gels consist of very long polymer chains \cite{hong2008theory,bouklas2015nonlinear,bertrand2016dynamics}.
Figure~\ref{fig:gel_osmo} shows the best fits for this equation to the osmotic pressure for each of our different hydrogels (dashline curves), obtained by fitting a constant value of $\chi$ to each data set.
These fits do not do a good job, which is not entirely surprising, as Flory-Huggins theory is known to make a number of assumptions that leads to poor quantitative comparison with experiments \cite{flory1942thermodynamics}. 

Instead, we again find good agreement with a simple power law: $\bar{\Pi}_\mathrm{mix}=B\phi^n$, as clearly shown by the fitted curves (solid curves) in Figure~\ref{fig:gel_osmo}, and the corresponding log-log plots in the inset. 
There are significant deviations in the PAM samples at high concentrations.
However, this occurs for $\phi>0.73$, where we expect the sample to become glassy.
Thus, it is not surprising to see a behavior change there, and we omit those points from the fitting.
The three PEGDA hydrogels have similar power-law exponents ($n=5.1, 4.6, 4.4$ for PEGDA 575, 700 and 1000 respectively), as highlighted by their similar slopes on the log-log plot.
However, this exponent is not universal:
the PAM 8.08\% and PAM 9.57\% hydrogels have significantly smaller exponents of $n=2.9$ and $n=3.2$, respectively.

\section{Relating $\Pi_\mathrm{mix}$ to gel microstructure}

Our observation of a power-law scaling for $\Pi_\mathrm{mix}$ may reveal information about the gel microstructure, mirroring similar results in polymer solutions \cite{rubinstein2003polymer}.
For example, semi-dilute polymer solutions typically obey Des Cloiseaux's theory \cite{des1975lagrangian}:
\begin{equation}
\frac{\Pi_\mathrm{mix}v_w}{k_bT}=A\phi^{3/(3-d)}.
    \label{eqn:powerlaw}
\end{equation}
Here, $A$ is a constant that depends on the polymer/solvent combination, and
$d$ is the fractal dimension of the polymer's conformation in microscopic `correlation blobs' in the solution.
This fractal nature comes from the self-avoiding random walks performed by the polymer chains in the blobs.
At scales significantly larger than the blobs, a solution is essentially uniform.
Thus, Equation~(\ref{eqn:powerlaw}) shows that measuring $\Pi_\mathrm{mix}(\phi)$ directly gives access to $d$. 

We hypothesize that Equation~(\ref{eqn:powerlaw}) also applies to many gels.
Like semi-dilute polymer solutions, gels are often thought of as being fractal in microscopic blobs below a certain correlation length, $\xi$ \cite{rubinstein2003polymer}.
For some gels -- like PAM with low crosslinker concentration -- this is because their polymer structure consists of long polymer chains that are loosely crosslinked together. Thus, their structure is similar to that of a solution of the same chains \cite{gombert2020hierarchical,cohen1992characterization}.
In gels made from short, network-forming oligomers -- like PEGDA -- the random gelation process should create a fractal network at the gel's mesh-scale \cite{muthukumar1986fractal,lazzari2016fractal}.
Thus, in general, the polymer arrangement in many gels can be described mathematically in the same way as a semi-dilute polymer solution: uniform at large scales, and fractal at small scales. Indeed, this structural similarity is often assumed when interpreting scattering experiments (\emph{e.g.} \cite{malo2015heterogeneity}).
We expect that $\Pi_\mathrm{mix}$ should only depend on polymer arrangement, so their structural similarity suggests that semi-dilute polymer solutions and fractal gels should have the same equation governing their osmotic pressure.
\emph{i.e.} Equation~(\ref{eqn:powerlaw}) should also apply to such gels.
This hypothesis can also be reached with a scaling argument, as shown in the Materials and Methods.

We use the assumption that Equation~(\ref{eqn:powerlaw}) applies to our gels to extract the apparent fractal dimension of their microstructures.
We equate Equation~(\ref{eqn:powerlaw}) to our empirical fitting $\bar{\Pi}_\mathrm{mix}=B\phi^n$ to obtain $A$ and $d=3-3/n$.
The resulting values, $A_o$ and $d_o$, are given in Table~\ref{tab1}.
For the PEGDA hydrogels, $d_o$ is in a narrow range between 2.32 and 2.41.
Indeed, when considering error bars on the measurement, these are consistent with a single value of $d_o = 2.37$ for the three gels.
However, this fractal dimension is not universal for all the gels.
$d_o$ takes a different, lower value for the polyacrylamide gels. 
$d_o=1.96$ and $2.06$ for low/high crosslinker-content respectively, consistent with a single value of $d_o=2.02$.
This is unsurprising, given the different microstructures expected in PAM and  PEGDA.
Interestingly, the value for PAM is close to $d=2$, which is seen when polymer chains in a theta solvent undergo ideal random walks.
For all the gels, $A_o$ is $O(1)$ -- as expected if the scaling arguments underlying the Equation~(\ref{eqn:powerlaw}) are correct. 

\begin{figure}[hpt]
\includegraphics[width=\columnwidth]{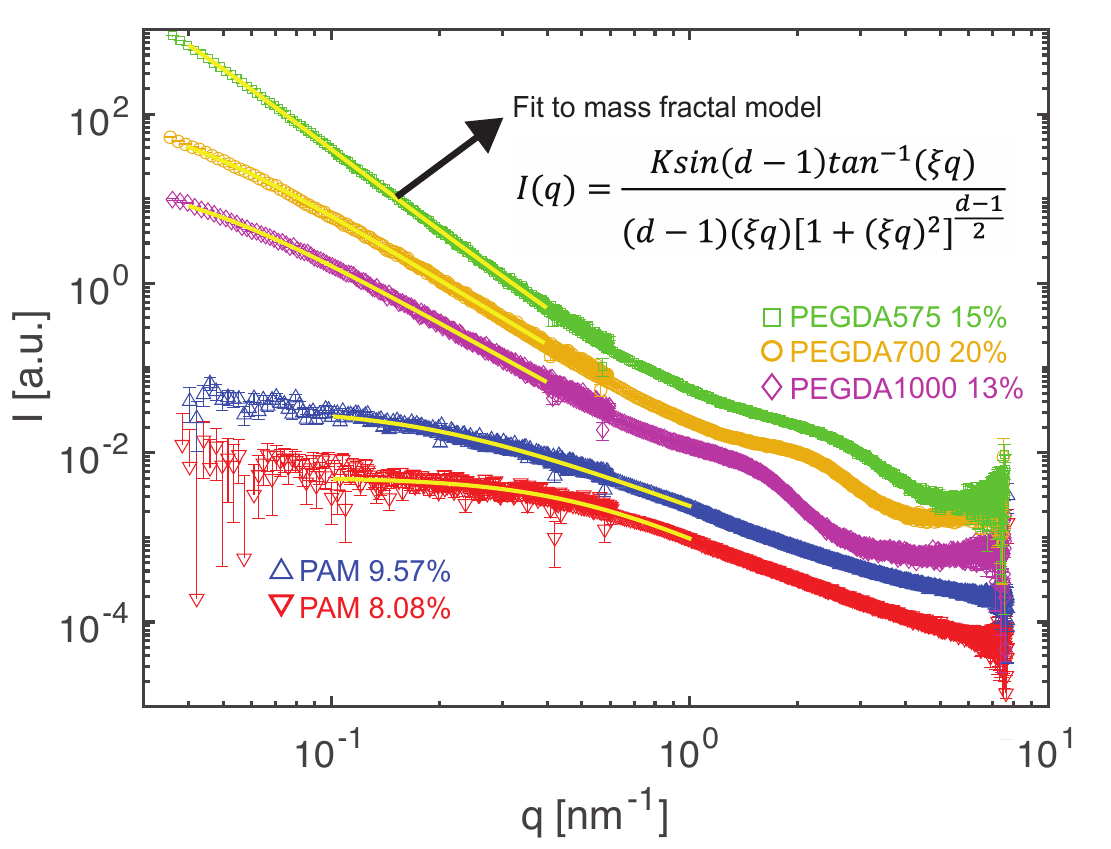}
\caption{Small-angle X-ray scattering profiles for the different hydrogel samples.
Curves are vertically shifted for ease of visualization.
Solid yellow curves show the curves obtained by fitting the low-$q$ data to Equation~(\ref{eqn:fractal}). For PEGDA hydrogels, we fit the range $0.04 \,\mathrm{nm}^{-1} <q< 0.4 \, \mathrm{nm}^{-1}$.
For PAM hydrogels, we fit the range $0.1 \,\mathrm{nm}^{-1} <q< 1 \, \mathrm{nm}^{-1}$ (lower $q$ is ignored, due to the large noise).
}
\label{fig:SAXS} 
\end{figure}

We test the extracted fractal dimensions, $d_o$ by performing small-angle X-ray scattering (SAXS) experiments on our hydrogels to detect and characterize fractal microstructures.
Figure~\ref{fig:SAXS} shows the scattered intensity, $I(q)$ for each gel, where $q$ is the magnitude of the scattering vector.
If a fractal microstructure exists with dimension $d$, we would expect it to give rise to a power-law decay in $I(q)\sim q^{-d}$ over the range of $q$ corresponding to the range of length-scales spanned by the fractal \cite{mildner1986small,seiffert2017scattering}.
At longer wavelengths -- \emph{i.e.} where $q\lesssim 2\pi/\xi$, the gel becomes uniform. Thus, we expect $I(q)$ to plateau at small $q$.
Indeed, this is exactly the behavior predicted by the equation describing the scattering of a fractal microstructure with correlation length $\xi$ \cite{mildner1986small}:
\begin{equation}    
I(q) = K\frac{\sin[(d - 1)\tan^{-1}(\xi q)]}{(d - 1)(\xi q) [ 1 + (\xi q)^2 ]^{\frac{d - 1}{2}}}.
\label{eqn:fractal}
\end{equation}
Here, $K$ is a constant pre-factor. 
Note that this is a generalization of scattering functions that describe semi-dilute polymer solutions, as equation (\ref{eqn:fractal}) reduces to the Lorentzian function when $d=2$ (the fractal dimension expected for polymer chains in a theta solvent).

The PEGDA hydrogels show rather convincing evidence of a fractal microstructure.
The PEGDA $I(q)$ curves in Figure~\ref{fig:SAXS} show power-law behavior for $q$ in the approximate range of $0.1~\mathrm{nm}^{-1}-1~\mathrm{nm}^{-1}$.
These $q$ values correspond to a range of length-scales that are somewhat bigger than the PEGDA oligomer size (the calculated radius of gyration, $R_g$, of each oligomer is given in Table~\ref{tab1}), where we expect to see the fractal behavior \cite{malo2015heterogeneity}. 
At even lower $q$, we also see $I(q)$ start to flatten out, indicative of the fact that we start to probe length-scales comparable to $\xi$.
Indeed, we can fit this whole, low-$q$ range of $I(q)$ with the fractal model in Equation~(\ref{eqn:fractal}). 
These fits are shown by the continuous curves in Figure~\ref{fig:SAXS}, with the corresponding fitting parameters ($d_s,\xi$) given in Table~\ref{tab1} (errors are 95\% confidence bounds for the fit).
Here, we choose to fit the range $q<0.4~\mathrm{nm}^{-1}$, as the scattering there should be independent of the form factor of the oligomers (see Supplement).
The fact that the fractal of dimension of PEGDA gels is always larger than $d=2$ is consistent with our expectation that this fractal structure is a crosslinked fractal mesh.

The PAM hydrogels can also be analyzed in the same way, although the evidence for a fractal microstructure is not so clear-cut.
The PAM curves in Figure~\ref{fig:SAXS} appear to show power-law behavior in a $q$ range in the approximate range of $0.8~\mathrm{nm}^{-1}-5~\mathrm{nm}^{-1}$.
This plateaus at lower $q$, and can be fitted with Equation~(\ref{eqn:fractal}) to obtain $\xi$ and $d_s$ (Table~\ref{tab1}).
However, the results must be viewed cautiously, as this fitting range is close to the atomic size, where we expect additional effects to control $I(q)$ -- such as contributions from the form factor of the Kuhn segments that make up the polymer (e.g. \cite{mcdowall2022using,beaucage1996small}).
To avoid these effects, and the noisy data at low $q$, we only fit the range $0.1~\mathrm{nm}^{-1}<q<1 ~\mathrm{nm}^{-1}$ (see Supplement for further details).
For the PAM gels, the measured fractal dimensions are consistent with that of polymer chains undergoing random walks (N.B. $d=2$ in a theta solvent, and $d=1.67$ in a good solvent \cite{beaucage1996small}). 
This is consistent with our expectation that PAM gels consist of loosely crosslinked polymer chains.

The extracted values of $d_s$ are close to measured values of $d_o$.
Indeed, for the two higher molecular weight PEGDA hydrogels, and the PAM 8.08\%, there is less than 5\% error.
For the other two hydrogels, the error is less than 20\%.
This is certainly reasonable agreement, given uncertainties in measuring fractal properties, which largely arise due to the limited $q$-range over which fractal behavior appears in $I(q)$.
Indeed, the observed agreement is surprisingly good, as we only expect agreement when there is a clear fractal in the hydrogel at all microscopic scales smaller than some length scale $\xi$ -- something which is not entirely obvious in all our samples.

To summarize, the fact that we see reasonable agreement between values of $d$ -- measured with two completely different techniques -- provides encouraging evidence for a relationship between a hydrogel's microstructure and osmotic pressure.
A true confirmation will require additional validation across a range of different materials.

\section{A simplified hydrogel constitutive model}

Our results have direct implications for modeling hydrogel mechanical behavior, as they motivate a simple constitutive model.
In particular, returning to Equation~(\ref{eqn:eff_stress}), we can insert our power-law expression for $\Pi_\mathrm{mix}(\phi)$ to obtain a new description of hydrogel behavior.
To complete this, we use one of the simplest descriptions of a polymer gel -- the phantom-network model \cite{james1953statistical,rubinstein2003polymer} -- which gives the principal stresses in the elastic network as
\begin{equation}
    \sigma_{i}^\mathrm{el}=\rho_c k_b T \frac{(\lambda_0\lambda_i)^2}{\lambda_0^3\lambda_1 \lambda_2 \lambda_3}.
    \label{eqn:sigma_el}
\end{equation}

Here, $\lambda_i$ are the principal stretches relative to the equilibrium swollen state polymer state, and $\lambda_0$ is the stretch from the dry state to the equilibrium swollen state. $\rho_c$ is a constant that is inversely proportional to the volume of a polymer chain between crosslinks.
With this, Equation~(\ref{eqn:eff_stress}) yields an expression for the principal stresses in the hydrogel as:
\begin{equation}
    \sigma_i=\frac{\rho_c k_b T}{\lambda_0} \frac{\lambda_i^2}{\lambda_1 \lambda_2 \lambda_3}-A\frac{k_bT}{v_w}\phi^n-\frac{\Delta\mu}{v_w}.
    \label{eqn:constit}
\end{equation}

We apply this constitutive model to our drained compression results in Figure~\ref{fig:stress-strain} by setting $\sigma_3=\sigma_{zz}=-P_\mathrm{ext}$ and $\Delta\mu=0$.
The corresponding stretches are $\lambda_3=\lambda=1/(\phi \lambda_0^3)$, and $\lambda_1=\lambda_2=1$.
Then, Equation~(\ref{eqn:constit}) becomes:
\begin{equation}    P_\mathrm{ext}=\frac{Ak_bT}{v_w}\phi^n-\frac{\rho_c k_bT \lambda}{\lambda_0}.
\end{equation}

In our experiment, $\lambda=1$ at 0~\textdegree{}C, while $P_\mathrm{ext}=0$.
Inserting this constraint allows us to eliminate the constant $\rho_c$:
\begin{equation}
    P_\mathrm{ext}=\frac{A k_bT}{v_w \lambda_0^{3n}}\left(\frac{1}{\lambda^n}-\lambda\right).
\end{equation}
This is exactly the same as the empirical fit that we found earlier (Equation \ref{eqn:emp_fit}).

Beyond our own data, we can apply this model to predict the equilibrium hydrogel swelling, $\lambda_0$.
For a fully-swollen gel in equilibrium with water, $\sigma_{ij}=0$, $\phi=\lambda_0^{-3}$, $\lambda_i=1$, and $\Delta\mu=0$.
Then, Equation~(\ref{eqn:constit}) becomes
\begin{equation}
    0=-\frac{Ak_bT}{v_w}\frac{1}{\lambda_0^{3n}}+\frac{\rho_c k_bT}{\lambda_0},
\end{equation}
so that the sample stretch between dry and equilibrium swelling is given by
\begin{equation}
    \lambda_0=\left(\frac{A}{\rho_c v_w}\right)^{\frac{1}{3n-1}}=\left(\frac{A}{\rho_c v_w}\right)^{\frac{3-d}{6+d}}.
\end{equation}
Here, the last equality assumes that $n$ is determined by a fractal structure of the hydrogel.
This shows how, the gel is expected to swell more, the smaller $d$ is.
Like the more complex Flory-Rehner theory \cite{flory1943statistical}, this equation relates the equilibrium swelling of a hydrogel to various material properties ($A$, $\rho_c$ and $d$, in particular). 
Thus, it could be used to extract information about these values from simple swelling experiments.

\section{Conclusions}

In conclusion, we have shown that freezing tests can be used to characterize hydrogel behavior over a wide swelling range.
In particular, these allow measurement of both a hydrogel's drained mechanical response under compression and its mixing osmotic pressure.
Our technique brings several advantages.
Firstly, it allows the measurement of hydrogel properties over a large compressive range.
For example, we could apply up to 25~MPa of pressure to samples in gel compression tests.
By way of comparison, previous works that used mechanical compression could apply maximum pressures that were over two orders of magnitude smaller \cite{li2012experimental,bhattacharyya2020hydrogel}.
Secondly, our technique permits the use of small samples.
Here, we used $\sim$100~\textmu{}L per experiment: much smaller than typical volumes used for mechanical compression tests.
Finally, our technique is much faster than alternative techniques.
The length of experiments is predominantly set by the time required for water to flow out of/into a gel so that it can reach thermodynamic equilibrium.
This time scales as $L^2/D$, where $L$ is the smallest dimension of the sample, and $D$ is the poroelastic diffusivity \cite{hu2012viscoelasticity}.
Due to this quadratic dependence upon $L$, thin samples will equilibrate much faster than thick ones.
Thus, while our $O$(100~\textmu{}m-thick) samples take only a few minutes to equilibrate, bulk samples used in other approaches take orders of magnitude longer.
In terms of limitations, there are two points worth bearing in mind:
this technique can only be used near the solvent's freezing temperature, and it can not be used with thermoresponsive materials, where $\Pi_\mathrm{mix}$ depends significantly upon $T$ (see Supplement for further details) \cite{madden2022comparison}.
These are identical to the limitations that arise in freezing-point osmometry.

Because we are able to measure properties over such a wide range of compressions, our technique permits new insights into hydrogel mechanical and osmotic behavior.
For example, the osmotic pressure of our gels has a simple, power-law dependence on polymer volume fraction, rather than following the predictions of classical polymer solution theory.
This casts doubt over the use of the wide range of constitutive models for polymer gels that are based upon this classical theory (\emph{e.g.} \cite{hong2008theory,kim2022polyacrylamide,bertrand2016dynamics,pan2022constitutive}).
Instead, our results motivate a modified constitutive model for gels.
In the future, it will be interesting to apply this model to hydrogel processes such as free-swelling, desiccation, fracture, volumetric phase transitions, and phase separation \cite{bertrand2016dynamics,mao2018theory,baumberger2006solvent,shibayama2005volume,fernandez2024elastic}.
Furthermore, our results suggest that gel osmotic pressure may be controlled by the form of the gel's microstructure.
In particular, we propose that a gel with a fractal microstructure should have a power-law dependence of osmotic pressure upon polymer content, with an exponent that is a simple function of the fractal dimension.
When we compare osmotic pressure measurements with fractal dimensions measured via scattering, we see that our results are consistent with this hypothesis.
However, further testing is needed for a complete validation.

There are several directions in which GelFrO can be extended.
In particular, it will be important to adapt the technique to characterize polyelectrolyte gels and gels containing solutes. 
These are more complex than our charge-neutral gels, as ions/solutes result in an increased gel osmotic pressure \cite{quesada2011gel}.
A separate direction could be to use free hydrogel samples (not attached to a rigid base) that can shrink isotropically during freezing.
Such tests are equivalent to isotropic, drained compression tests. Combining such results with uniaxial compression results, like those presented here, will allow the accurate fitting of constitutive models to data from multiple different deformation modes: a necessity for accurate model fitting \cite{ricker2023systematic}.


\section*{Materials and Methods}

\subsection{Experiments}

We prepare hydrogels in temporary sample cells.
These are made with two 150~\textmu{}m-thick, glass coverslips, separated by 150~\textmu{}m spacers. 
The bottom coverslip is silanized with 3-(trimethoxysilyl)propyl acrylate to promote hydrogel bonding \cite{yang2024dehydration}.
We fill the cells with hydrogel precursor solution containing a photoinitiator, and expose them to UV light (wavelength: 365~nm) for 1 hour at a power density of (0.39~$\pm$~0.07)~mW/cm$^2$.
The resulting polymerized gels are then left overnight at room temperature in a humid container before imaging.

To make PEGDA precursor solutions, we mix de-ionized water (18.2~M$\Omega$cm) with PEGDA of different molecular weights (575, 700 and 1000~g/mol from Sigma-Aldrich at respective concentrations of 15, 20 and 13 vol\%), 0.1 vol\% of 2-hydroxy-2-methylpropiophenone photoinitiator (Tokyo Chemical Industry), and 0.05 vol\% of fluorescent nanoparticles (200~nm diameter, red, carboxylate-modified Fluospheres, Thermo Fisher Scientific).
To make the two different polyacrylamide precursor solutions, we mix deionized water with acrylamide monomer (Sigma-Aldrich), N, N'-Methylenebisacrylamide crosslinker (Sigma-Aldrich), 0.1 vol\% of photoinitiator, and 0.05 vol\% of fluorescent nanoparticles.
The first polyacrylamide solution has 8.01 vol\% of monomer and 0.07 vol\% of crosslinker.
The second solution has 9.08 vol\% of monomer and 0.39 vol\% of crosslinker.
All precursor solutions are thoroughly mixed and sonicated for 10 mins before use.

In our experiments, we measure the equilibrium thickness of our hydrogels under different conditions with a confocal microscope.
We take 3-dimensional (3-D) image stacks of fluorescent particles dispersed in the gels with a 20x air objective (Numerical aperture 0.17) on a confocal microscope (Nikon Ti2 Eclipse) equipped with a Spinning Disk Confocal system (561~nm laser, 3i).
We first image as-prepared hydrogels at various marked positions in the temporary sample cells (away from the cell edges).
We then convert the temporary cells into freezing cells.
We remove the top glass coverslip from the temporary cells to leave the bottom coverslip with its attached hydrogel coating.
We then add 450~\textmu{}m spacers with a new top coverslip, and fill the resulting gap above the hydrogel with de-ionized water (Figure~\ref{fig:schematic} top).
We then place the freezing cells in the freezing stage (Figure~\ref{fig:schematic}) and mount them on the confocal microscope.
Next, we reduce the temperature of the sample from room temperature to 0~\textdegree{}C in a step-wise fashion, while regularly imaging the hydrogel.
Finally, we nucleate ice in the cell (Figure \ref{fig:schematic} top), and continue to step-wise reduce the temperature below 0~\textdegree{}C, while imaging the hydrogel.

To form ice in the freezing cell, we first cool one side of the sample slightly below 0~\textdegree{}C.
This is achieved by independently tuning the temperatures of the copper blocks in our freezing stage.
Then, we nucleate ice by touching the cold side of the sample with a cotton swab that has been dipped in liquid nitrogen.
After nucleation, we slowly decrease the temperature of the warm side of the cell (at 0.1~\textdegree{}C/min) until there is a uniform, subzero temperature across the whole sample, and proceed with the freezing experiment.

We measure relative thickness changes in the hydrogel by tracking displacements of fluorescent nanoparticle tracers dispersed throughout in the gel.
We first locate 3-D tracer positions in confocal image stacks using a code based on Matlab's regionprops3 function.
We then calculate tracer displacements by tracking changes in these positions between time-points, using a large-displacement tracking code \cite{kim2021measuring}.
Finally, we use these displacements to calculate the relative (uniform) vertical shrinkage, $h/h_0$ of the hydrogel layer.
This analysis involves the use of a small correction accounting for the hydrogel's refractive index (e.g. see the supplement of  \cite{style2014traction}, and the Supplement of this work).

Small-strain, drained moduli for the hydrogels are measured via long-time poroelastic indentation experiments \cite{hu2010using}.
We indent bulk hydrogel samples (1.3~cm thickness, 9~cm diameter) with a 10~mm radius, spherical indenter attached to a texture analyzer with a 500~g load cell (TA.XT Plus, Stable Microsystems).
The results are analyzed following \cite{hu2010using} to give the drained shear modulus, and drained Poisson ratio, from which $E_d$ is calculated. For details see the Supplement.

We perform SAXS measurements using a Xenocs SAXS instrument, equipped with a Cu radiation ($\lambda =$ 0.154~nm) source. 
We make samples in quartz capillaries with an outer diameter of 1 mm and a wall thickness of $<$\ 0.1~mm. 
For each sample, we collect data using two different sample-to-detector distances, $L_{\rm sd}$. Firstly, we record data with $L_{\rm sd}=350$ mm, and a beam size of 0.7~mm $\times$ 0.7~mm.
In this case, the magnitude of the scattering vector, $q$ varies between 0.13~nm$^{-1}$ and 7.3~nm$^{-1}$.
Subsequently, we record data with $L_{\rm sd}=1650$ mm, and a beam size of 0.25~mm $\times$ 0.25~mm.
Then, $q$ varies between 0.025~nm$^{-1}$ and 1.3~nm$^{-1}$.
For both cases, we record for 900 s to achieve an adequate signal-to-noise ratio.
Each test is repeated 5 times for each sample, and the results are combined and averaged.
Finally, we perform a background subtraction, by subtracting the scattering intensity from a deionized-water-filled quartz capillary -- yielding the scattered intensity, $I(q)$.

\subsection{$\Delta \mu$ for hydrogels in equilibrium with ice}

To derive Equation~(\ref{eq:cryosuc_versus_gel_balance}), we need an expression for the difference between the chemical potential of water in a hydrogel that is in equilibrium with ice at ($T,P_a$), and the chemical potential of water at ($T,P_a$), where $P_a$ is atmospheric pressure, and $T<T_m$:
\begin{equation}
    \Delta\mu=\mu_i(P_a,T)-\mu_w(P_a,T).
\end{equation}
Here, $\mu_w$ and $\mu_i$ are the chemical potentials of water and ice respectively, and we have used the fact that water in equilibrium with ice has the same chemical potential as the ice.

To evaluate this, we Taylor-expand this expression about $(P_a,T_m)$, and recall that $\mu_i(P_a,T_m)=\mu_w(P_a,T_m)$ (because water and ice are in equilibrium at ($P_a,T_m$)) to find that
\begin{equation}
    \Delta\mu=(T-T_m)\frac{\partial\mu_i}{\partial T}-(T-T_m)\frac{\partial\mu_w}{\partial T}.
\end{equation}
For a simple, one-component material, $\mu=g m$, where $g$ is the specific Gibbs free energy, and $m$ is the molecular mass. Further, $dg=-sdT+(1/\rho)dP$, where $\rho$ is the material's density, and $s$ is the specific entropy \cite{style2023generalized}.
Thus, $\partial \mu/\partial T=m\partial g/\partial T=-ms$, and we obtain that
\begin{equation}
    \Delta \mu=m(T-T_m)\left(s_w-s_i\right)=-m L\frac{(T_m-T)}{T_m}.
\end{equation}
In the last equality, we have used the identity $L=(s_w-s_i)T_m$.
Finally, we note that $m/v_w=\rho_w$, so 
\begin{equation}
    \frac{\Delta \mu}{v_w}=-\rho_w L\frac{(T_m-T)}{T_m},
\end{equation}
which is the result used in the main text.

\subsection{The mixing osmotic pressure of a fractal gel}

From purely dimensional arguments, $\Pi_\mathrm{mix}$ for a fractal gel must take the form
\begin{equation}
    \Pi_\mathrm{mix}=k_bT c f(\phi,N),
\end{equation}
where $f$ is an unknown, dimensionless function, $\phi$ is the volume fraction of polymer, $N$ is the degree of polymerization of the polymer, and $c=\phi/b^3$ is the number density of monomers in the gel, where $b$ is the monomer size.
While $f\propto N^{-1}$ for short chains in Van't Hoff's regime \cite{DoiPolymPhys}, it should become independent of polymerization degree for large $N$ \cite{rubinstein2003polymer}, so $f=f(\phi)$.
This is essentially a generalized virial expansion for $\Pi_\mathrm{mix}$. 

We now use the scaling hypothesis \cite{de1979scaling} that says that the expression for $\Pi_\mathrm{mix}$ should remain unchanged, if we replace our monomer with a `supermonomer' consisting of an arbitrary number $\lambda$ of monomers.
In this case, the supermonomer concentration is $c'=c/\lambda$, and the size of a supermonomer is $b'=\lambda^\nu b$.
The latter expression comes from the fractal nature of the network, with fractal dimension $d=1/\nu$.
Thus,
\begin{equation}
    \Pi_\mathrm{mix}=k_bT c f(c b^3)=k_bT c' f(c' b'^3).
\end{equation}
Finally, we use the common power-law ansatz, $f(\phi)\propto\phi^\alpha$ \cite{DoiPolymPhys}.
Thus,
\begin{eqnarray}
    c f(c b^3)&=& A' c^{\alpha+1} b^{3\alpha}=A' c'^{\alpha+1} b'^{3\alpha}
    \nonumber \\
    &=&A' c^{\alpha+1} b^{3\alpha} \lambda^{\alpha(3\nu-1)-1},
\end{eqnarray}
where $A'$ is a dimensionless constant.
The only way for this to be insensitive to the choice of $\lambda$, is if $\alpha=(3\nu-1)^{-1}$.
Thus we find that
\begin{equation}
    \Pi_\mathrm{mix}=A'k_bTc \phi^{1/(3\nu-1)}=A'\frac{k_bT}{b^3}\phi^{3/(3-d)}.
\end{equation}
Defining $A=A'v_w/b^3$, we get the result \eqref{eqn:powerlaw} in the main text.

\begin{acknowledgments}

We thank Shaohua Yang, Dr. Charlotta Lorenz, Dr. Viviane L\"{u}tz and Dr. Alan Rempel for their helpful discussions. 
We thank Dr. Thomas Weber for helping with SAXS measurements. 
We acknowledge support from the Swiss National Science Foundation (200021-212066, PZ00P2186041).

\end{acknowledgments}

\appendix



%

\end{document}